\documentclass[12pt]{article}

\usepackage{amsxtra,amssymb,amsthm,amsmath,latexsym}

\theoremstyle{plain}

\newtheorem{theorem}{Theorem}

\newtheorem{question}{Question}

\def\R{{\mathbb R}}

\def\oH{\buildrel\circ\over H}
\def\oH1{\buildrel\circ\over H\kern-.02in{}^1}
\def\qed{{\hfill $\Box$}}

\def\bysame{\leavevmode\hbox to3em{\hrulefill}\thinspace}

\begin{document}


\title{
Inequalities for the derivatives
   \thanks{key words:  stable numerical differentiation, inequalities
   for the derivatives }
   \thanks{Math subject classification: 65D25, 65M10 }
}

\author{
A.G. Ramm\\
 Mathematics Department, Kansas State University, \\
 Manhattan, KS 66506-2602, USA\\
{\it \small ramm@math.ksu.edu}}

\date{}
\maketitle\thispagestyle{empty}

\begin{abstract}

The following question is studied and answered: 

Is it possible 
to stably approximate $f^\prime$ if one knows:

1) $f_\delta\in L^\infty(\R)$ such that $\|f-f_\delta\|<\delta$,

 and

2) $f\in C^\infty(\R)$, $\|f\|+\|f^\prime\|\leq c$?

Here $\|f\|:=\sup_{x\in\R} |f(x)|$ and $c>0$ is a given constant.
By a stable approximation one means
$\|L_\delta f_\delta-f^\prime\|\leq \eta(\delta)\to 0$
as $\delta\to 0$. By $L_\delta f_\delta$ one denotes an estimate
of $f^\prime$.
The basic result of this paper is the inequality for
$\|L_\delta f_\delta-f^\prime\|$,
a proof of the impossibility to approximate stably $f^\prime$
given the above data 1) and 2), and a derivation of the inequality
$\eta(\delta)\leq c\delta^{\frac a{1+a}}$ if 2) is replaced by
$\|f\|_{1+a}\leq m_{1+a}$, $0<a\leq 1$. An explicit formula for
the estimate $L_\delta f_\delta$ is given.

\end{abstract}


\section{Introduction}
The classical problem of theoretical and computational mathematics is
the problem of estimation of the derivative $f^\prime$ of a
function from various data.

Inequalities between the derivatives are known (Landau-Hadamard,
Kolmogorov \cite{HLP}-\cite{L}, \cite{R1}), for example:

\begin{equation}
  m_k\leq c_{nk}m_0^{\frac{n-k}{n}} m_n^{\frac{k}{n}},
  \tag{1.1} \end{equation}
where 
$$m_k:=\|f^{(k)}\|:=\sup_{x\in I} |f^{(k)}(x)|, \,\, I=\R,
$$
and $c_{nk}$ are some constants. In particular, if $I=\R$, then
\begin{equation}
  m_1\leq \sqrt{2m_0m_2},
  \tag{1.2} \end{equation}
if $I=(0,\infty)$, then
\begin{equation}
  m_1\leq 2\sqrt{m_0m_2},
  \tag{1.3} \end{equation}
if $I=(0,h)$, $h\geq 2\sqrt{\frac{m_0}{m_2}}$, then  (1.3) holds,
if $I=(0,h)$, $h < 2\sqrt{\frac{m_0}{m_2}}$, then
\begin{equation}\
  m_1\leq \frac{2}{h} m_0+\frac{h}{2}m_2.
  \tag{1.4} \end{equation}
These inequalities can be found in \cite{HLP}-\cite{L}.

In pratice the following problem is of great interest.
Suppose that $f(x)\in C^\infty(\R)$ is unknown, but
one knows $m_j$, $j=0,1,2$, and one knows $f_\delta\in L^\infty(\R)$
such that
\begin{equation}
  \|f_\delta-f\|\leq\delta.
  \tag{1.5} \end{equation}
Can one estimate $f^\prime(x)$ stably? In other words,
can one find an operator $L_\delta$ such that
\begin{equation}
  \|L_\delta f_\delta-f^\prime\|\leq \eta(\delta)\to 0
  \qquad\hbox{as}\quad \delta\to 0.
  \tag{1.6} \end{equation}
The operator $L_\delta$ can be linear or nonlinear, in general.

This problem was investigated in \cite{R2}, where
it was proved that the operator
\begin{equation}
  L_\delta f_\delta:=
  \frac{f_\delta(x+h(\delta))-f_\delta(x-h(\delta))}{2h(\delta)},
  \qquad h(\delta):=\sqrt{\frac{2\delta}{m_2}}
  \tag{1.7} \end{equation}
yields the estimate:
\begin{equation}
  \|L_\delta f_\delta-f^\prime\|
  \leq \varepsilon(\delta):=\sqrt{2m_2\delta},
  \tag{1.8} \end{equation}
under the assumptions $m_2<\infty$ and (1.5).

Inequality (1.8) is quite convenient practically.
The original result of \cite{R2} was the first of its kind and
generated many papers in which the choice of the
discretization parameter was used for a stable solution of
various ill-posed problems, in particular stable
differentiation of random functions and applications
in electrical engineering
(see \cite{MR}-\cite{R6} and references therein).

In \cite[pp.82-84]{R1} one can find a proof
of the following interesting fact:
{\it among all linear and nonlinear operators $T$, the
operator $L_\delta$, defined in (1.7), gives the best possible estimate
of $f^\prime$ on the class of all $f\in{\mathcal K}(\delta,m_2)$.}
Here
\begin{equation}
  \mathcal K(\delta,m_j):=\{f:f\in C^j(\R),
  \quad m_j<\infty, \quad \|f-f_\delta\|\leq\delta\}.
  \tag{1.9} \end{equation}
In other words, the following inequality holds \cite[p.82]{R1}:
\begin{equation}
  \mathop{\inf}_{T} \sup_{f\in\mathcal K(\delta,m_2)}
  \|T f_\delta-f^\prime\|\geq\varepsilon(\delta):=\sqrt{2m_2\delta},
  \tag{1.10} \end{equation}
where $T$ runs through the set of all linear and nonlinear operators
$T:L^\infty(\R)\to L^\infty(\R)$. 

In this paper we investigate and answer the following questions:

\begin{question} 
{Given $f_\delta\in L^\infty(\R)$ such that (1.5) holds,
and a number $m_j$, $\|f^{(j)}\|\leq m_j$, $f\in C^\infty(\R)$,
$j=0,1$, can one estimate stably $f^\prime$?}
\end{question}

In other words, does there exist an operator  $T$ such that
\begin{equation}
  \sup_{f\in\mathcal K(\delta,m_j)}
  \|Tf_\delta-f^\prime\|\leq\eta(\delta)\to 0
  \qquad\hbox{as}\quad \delta\to 0,
  \tag{1.11} \end{equation}
where $j=0$ or $j=1$?

\begin{question} 
It is similar to Question 1 but now it is assumed that $j=1+a>1$:
\begin{equation}
  \|f^{(1+a)}\|:=m_{1+a}<\infty, \quad 0<a\leq 1,
  \tag{1.12} \end{equation}
where $\|f^{(1+a)}\|:=\|f^{\prime^{(a)}}\|$, and
\begin{equation}
  \|g^{(a)}\|:=\sup_{x,y\in\R}
  \frac{|g(x)-g(y)|}{|x-y|^a} +\|g\|,
  \qquad 0<a\leq 1.
  \tag{1.13} \end{equation}
\end{question}

The basic results of this paper are summarized in Theorem 1.

\begin{theorem}
There does not exist an operator $T$ such that inequality (1.11)
holds for $j=0$ or for $j=1$. There exists such an operator if $j>1$.
\end{theorem}

In the proof of Theorem 1 an explicit formula is given for $T$
and an explicit inequality (2.8) is given for the error estimate.

In section 2 proofs are given. In the course of these proofs we derive
inequalities for the quantity
\begin{equation}
  \gamma_j:=\gamma_j(\delta):=
  \gamma_j(\delta,m_j):=\inf_T\sup_{f\in K(\delta,m_j)}
  \|Tf_\delta-f^\prime\|
  \tag{1.14} \end{equation}
In \cite{R7} the theory presented in this paper
is developed further and numerical examples of its applications
are given.

\section{Proof of Theorem 1}

Let $f_\delta(x)=0$, and consider $f_1(x):=-\frac{M}{2}x(x-2h)$,
$0\leq x\leq 2h$, and $f_1(x)$ is extended to the whole real axis
in such a way that
$\|f^{(j)}_1\|=\sup_{0\leq x\leq 2h}\|f^{(j)}_1\|$, $j=0,1,2,$
are preserved. It is known that such an extension is possible.
Let $f_2(x)=-f_1(x)$.
Denote $(Tf_\delta)(0):=(T0)(0):=b$.

Since
\begin{equation}
  \|Tf_\delta-f^\prime_1\|
  \geq \left| (Tf_\delta)(0)-f^\prime_1(0)\right|
  =|b-Mh|, \notag\end{equation}
and
\begin{equation}
  \|Tf_\delta-f^\prime_2\|\geq|b+Mh|,
  \notag\end{equation}
one has
\begin{equation}
  \gamma_j(\delta)\geq \inf_{b\in\R}
  \max \left\{|b-Mh|,|b+Mh|\right\}=Mh
  \tag{2.1} \end{equation}

Inequality (1.5) with $f_\delta(x)=0$ implies
\begin{equation}
  \sup_x |f_s(x)|=\frac{Mh^2}{2}\leq \delta,\qquad s=1,2.
  \tag{2.2} \end{equation}

Let us take $\frac{Mh^2}{2}=\delta$, then
\begin{equation}
  h=\sqrt{\frac{2\delta}{M}},\qquad Mh=\sqrt{2\delta M}.
  \tag{2.3} \end{equation}
If $j=0$, then (2.2) implies $m_0=\delta$.
Since $M$ can be chosen arbitrary for any $\delta>0$
and $m_0=\delta$,
inequality (2.1) with $j=0$ proves that estimate (1.11) is false
on the class
${\mathcal K}(\delta,m_0)$, and in fact $\gamma_0(\delta)\to\infty$
as $M\to\infty$.

This estimate is also false on the class ${\mathcal K}(\delta, m_1)$.
Indeed, for $f_1(x)$ and $f_2(x)$ one has
\begin{equation}
  m_1=\|f^\prime_1\|= \|f^\prime_2\|= \sup_{0\leq x\leq 2h} |M(x-h)|
  =Mh=\sqrt{2\delta M}.
  \tag{2.4} \end{equation}
If $m_1\leq c<\infty$, then one can find $M$ such that
$m_1=\sqrt{2\delta M}=c$, thus $Mh=c$, and by (2.1) one gets
\begin{equation}
  \gamma_1(\delta)\geq c>0,\qquad \delta\to 0,
  \tag{2.5} \end{equation}
so that (1.11) is false.

Let us assume now that (1.12) holds.
Take $Tf_\delta:=L_{\delta,h}f_\delta$,
where  $L_{\delta,h}f_\delta$ is defined as in (1.7) but $h$ replaces
$h(\delta)$.
One has, using the Lagrange formula,
\begin{equation}
  \begin{align}
  \|L_{\delta,h}f_\delta-f^\prime\|
    =&\|L_{\delta,h}(f_\delta-f)\|
    +\left\|L_{\delta,h}f-f^\prime \right\|
    \notag \\
  \leq & \frac{\delta}{h}
    +\left\|\frac{f(x+h)-f(x-h)-2hf^\prime(x)}{2h} \right\|
    \notag \\
  \leq &\frac{\delta}{h}
    +\left\|\frac{[f^\prime(y)
       -f^\prime(x)] h+[f^\prime(z)-f^\prime(x)]h}{2h} \right\|
    \notag \\
  \leq & \frac{\delta}{h} + m_{1+a}h^a:=\varepsilon_a(\delta,h).
    \tag{2.6}   \end{align}
    \notag \end{equation}
where $y$ and $z$ are 
the intermediate points in the Lagrange formula.

Minimizing the right-hand side of (2.6) with respect to $h\in(0,\infty)$
yields
\begin{equation}
  h_a(\delta)=\left(\frac{\delta}{am_{1+a}}\right)^{\frac{1}{1+a}},
  \qquad \varepsilon_a(\delta)=c_a\delta^{\frac{a}{1+a}},
  \qquad 0<a\leq 1,
  \tag{2.7} \end{equation}
where $c_a:=\left( am_{1+a} \right)^{\frac{1}{1+a}}
+\frac{m_{1+a}}{\left(am_{1+a}\right)^\frac{a}{1+a}}$.

From (2.6) and (2.7) the following inequality follows:
\begin{equation}
  \sup_{f\in {\mathcal K}(\delta,m_{1+a})}
  \|L_\delta f_\delta-f^\prime\|
  \leq c_a \delta^{\frac{a}{1+a}},
  \qquad 0<a\leq 1.
  \tag{2.8} \end{equation}

Theorem 1 is proved. \qed

\vfil\pagebreak

\end{document}